\documentclass[aps,prl,twocolumn]{revtex4-2}
\pdfoutput=1
\usepackage[caption=false]{subfig}
\usepackage{array}
\usepackage{slashed,bbold}
\usepackage{physics}
\usepackage{dsfont}
\usepackage{lipsum}
\usepackage{booktabs}

\usepackage{amsmath,amssymb,bm} 
\usepackage{graphicx}
\usepackage{soul}
\usepackage{xcolor}

\usepackage[papersize={8.5in,11in}]{geometry}

\usepackage{color}
\definecolor{darkblue}{rgb}{0.,0.,0.4}
\definecolor{darkred}{rgb}{0.5,0.,0.}
\definecolor{BlueViolet}{RGB}{138,43,226}
\definecolor{SkyBlue}{RGB}{30,144,255}
\definecolor{DarkGreen}{RGB}{0,100,0}
\usepackage[pdftex,colorlinks=true,linkcolor=darkblue,citecolor=blue,urlcolor=darkred]{hyperref}

\def \be{\begin{equation}}
\def \ee{\end{equation}}

\begin{document}
	
\title{Phase-Shifted Planar Hall and Magnetoresistive Responses in Weyl Semimetals }
\author{Sandip Bera$^1$ and Sudhansu S. Mandal$^{2}$ }
\affiliation{$^1$Department of Physics, University of Toronto, 60 St. George Street, Toronto, Ontario, Canada M5S 1A7	\\ 
	$^2$Department of Physics, Indian Institute of Technology, Kharagpur 721302, India
}

\date{\today}
	
\begin{abstract}
	
The planar Hall resistivity and magnetoresistivity of Weyl semimetals are conventionally expected to exhibit $\sin 2\phi$ and $\cos2\phi$ angular dependences, respectively, where $\phi$ is the angle between electric and magnetic fields. However, experiments reveal shifted extrema in the planar Hall signal and sign reversals in magnetoresistivity at $\phi < \pi/4$. Here, using a diagrammatic Kubo-formula approach, we identify an intrinsic 
quadratic magnetic-field contribution to the planar transport response that is absent in conventional semiclassical description. This contribution introduces an additional term proportional to $\cos 2 \phi$ and $\sin 2\phi$, respectively, in transverse and longitudinal conductivities.  
Consequently, both planar Hall and magnetoresistive responses acquire a phase-shifted form  $\sin 2(\phi+\phi_r)$ and  $\cos 2(\phi+\phi_r)$, respectively. 
The same phase shift extracted independently from longitudinal and transverse responses quantitatively describes available experimental data. Our results establish a microscopic origin of the anomalous angular dependence observed in Weyl semimetals.

\end{abstract}

\maketitle

Topological quantum materials host electronic states with unconventional responses arising from their nontrivial band topology.  Weyl semimetals (WSMs) \cite{PhysRevB.83.205101,PhysRevB.84.235126,PhysRevB.85.165110,Sun2012,PhysRevA.85.033640,PhysRevLett.108.046602,Delplace_2012} are one such class of materials where pairs of linearly dispersing band crossing occur at isolated points in momentum space.These points are known as Weyl nodes and the quasiparticles near these nodes behave as Weyl fermions \cite{Weyl1929}. A pair of Weyl nodes with opposite chiralities appear as separated in either (both) momentum or (and) energy, depending on the broken time reversal and parity symmetries in the underlying crystal.

A defining feature of Weyl fermions is the chiral anomaly \cite{Adler1969,BellJackiw1969}: parallel electric and magnetic fields pump charge between fermions with opposite chiralities.   This anomaly produces anisotropic magnetotransport \cite{Son2013} in Weyl semimetals. The chiral anomaly provides also a microscopic basis for the planar Hall effect in Weyl semimetals when electric field $\mathbf{E}$ and  magnetic field $\mathbf{B}$ are coplanar. It manifests as anisotropic magneto-conductivity (MC) and planar Hall conductivity (PHC) that may be expressed \cite{Burkov2017,Nandy2017} through planar current density as $\mathbf{J} = \sigma_0 \mathbf{E} + \alpha_1 (\mathbf{E} \cdot \mathbf{B})\mathbf{B}$, with MC, $\sigma_{{\rm mc}} = \sigma_{xx}(B)-\sigma_{xx}(0)= \alpha_1 B^2 \cos^2\phi$, and PHC, $\sigma_{{\rm phc}}=\sigma_{yx}(B)=\alpha_1 B^2 \sin\phi\cos\phi$, where $\phi$ is the angle between $\mathbf{E}$ and $\mathbf{B}$. Here $\sigma_0$ is the Drude conductivity and the coefficient $\alpha_1$ is proportional to the relaxation time $\tau$ in the Boltzmann transport theory. This conventional description treats that the magnetic-field-induced current is entirely parallel to the magnetic field through the term $(\mathbf{E} \cdot \mathbf{B})\mathbf{B}$. However, in a crystal with a preferred Weyl-node separation axis, additional symmetry-allowed terms can contribute to the quadratic magnetic response.

The transport measurements have indeed demonstrated \cite{Zhang2019,Wu2018,Zhang2016,Veyrat2024,Yang2019,Niemann2017,Li2018magnetoresistance,yang2015,Huang2015,Li2016,Arnold2016} that  PHC in WSMs is proportional to square of the applied low magnetic field and is anisotropic as proposed \cite{ Burkov2017,Nandy2017}. However, in contrary to the results  \cite{ Burkov2017,Nandy2017} of Boltzmann transport theory that counts the contribution of the electrons only at the Fermi surface, the extremum of $\sigma_{{\rm phc}}$ does not occur at $\phi = \pi/4$ and even more surprisingly $\sigma_{{\rm mc}}$ changes \cite{Niemann2017,Li2018magnetoresistance,yang2015,Huang2015,Li2016,Arnold2016} sign within the range $0 < \phi <\pi/4$. To this end, the quadratic-order magnetic field correction to the current density allowed by symmetry can be written as
\begin{eqnarray}
	\mathbf{J} &=& \sigma_0 \mathbf{E} + \alpha_1 (\mathbf{E} \cdot \mathbf{B})\mathbf{B} +
	\alpha_2 \left(\mathbf{E}  \cdot (\hat{z} \times \mathbf{B})\right)\mathbf{B}
	 \nonumber \\
	&+&  \alpha_3 (\mathbf{E} \cdot \mathbf{B})(\hat{z} \times \mathbf{B} ) +\alpha_4\left( \mathbf{E}  \cdot (\hat{z} \times \mathbf{B})\right) (\hat{z} \times \mathbf{B}  )\, .
\end{eqnarray}
The additional structures arise because $\mathbf{E} \cdot \mathbf{B}$ and $\mathbf{E} \cdot (\hat{z} \times \mathbf{B})$  are allowed scalars once a Weyl-node separation axis breaks rotational symmetry, and simultaneously current can flow both along $\bm{B}$ and its transverse direction $\hat{z}\times \bm{B}$. 
 This leads to 
\begin{eqnarray}
	\sigma_{{\rm phc}} = B^2( \alpha_1  \sin2\phi + \alpha_2 \cos2\phi)= \sigma_r \sin 2(\phi + \phi_r) \, \,\,\,  \label{sigma_phc} \\
	\sigma_{{\rm mc}}  = B^2(\alpha_1 \cos2\phi -
	 \alpha_2 \sin2\phi )= \sigma_r  \cos 2(\phi + \phi_r)\, \,\,\, \label{sigma_mc}
\end{eqnarray}
with $\sigma_r =B^2 (\alpha_1^2+\alpha_2^2)^{1/2}$ and   $\phi_r = \frac{1}{2}\arctan (\alpha_2/\alpha_1)$, if $\alpha_4=-\alpha_1$ and $\alpha_3 =  \alpha_2$. Clearly, the extremum of $\sigma_{\rm phc}$ will shift from $\phi = \pi/4$ by $\phi_r$ and $\sigma_{\rm mc}$ will change sign at $\phi = \pi/4-\phi_r$.

We next microscopically evaluate these symmetry-allowed terms using a diagrammatic \cite{Altshuler1980,Varlamov1991} Kubo formalism that includes both Fermi-surface and Fermi-sea contributions. We indeed find  $\sigma_{{\rm phc}}$ and $\sigma_{{\rm mc}}$, as formally expressed in Eqs.~(2) and (3), in Kubo formalism.  As $\sigma_{xx}$ is dominated by $\sigma_0$, diagonal and Hall resistivities approximately become  $\rho_{xx}  \approx 1/\sigma_{xx}$ and $\rho_{yx} \approx - \sigma_{yx}/\sigma_0^2$. Therefore
 \begin{eqnarray} \label{Resistivity}
 	\rho_{xx} &=& \rho_0 - \rho_r \cos2(\phi +\phi_r)  \label{Resistivity1} \\
 	\rho_{yx} & =&  -\rho_r \sin2(\phi +\phi_r) \label{Resistivity2}
 \end{eqnarray}	
where $\rho_0 = 1/\sigma_0$ and $\rho_r = \sigma_r /\sigma_0^2$ is the resistivity amplitude of the periodic variation. We first test the shifted angular dependence against experimental data and extract the phase shift and amplitude, and subsequently derive the coefficients $\alpha_1$ and $\alpha_2$ microscopically within a diagrammatic Kubo formalism.

Equation \eqref{Resistivity1} predicts that the magnetoresistive correction  $\Delta \rho_{xx} = \rho_{xx}(B) - \rho_0$ changes  sign at $\phi = \pi/4 -\phi_r$. Reported measurements in  NbP \cite{Niemann2017}, NbAs \cite{yang2015}, TaAs \cite{Huang2015}, Cd$_3$As$_2$ \cite{Li2018magnetoresistance,Wu2018}, and TaP  \cite{Arnold2016} show sign changes at angles corresponding to finite phase shifts $\phi_r$.

\begin{figure}[h]
	\begin{center}
		\includegraphics[scale=0.65]{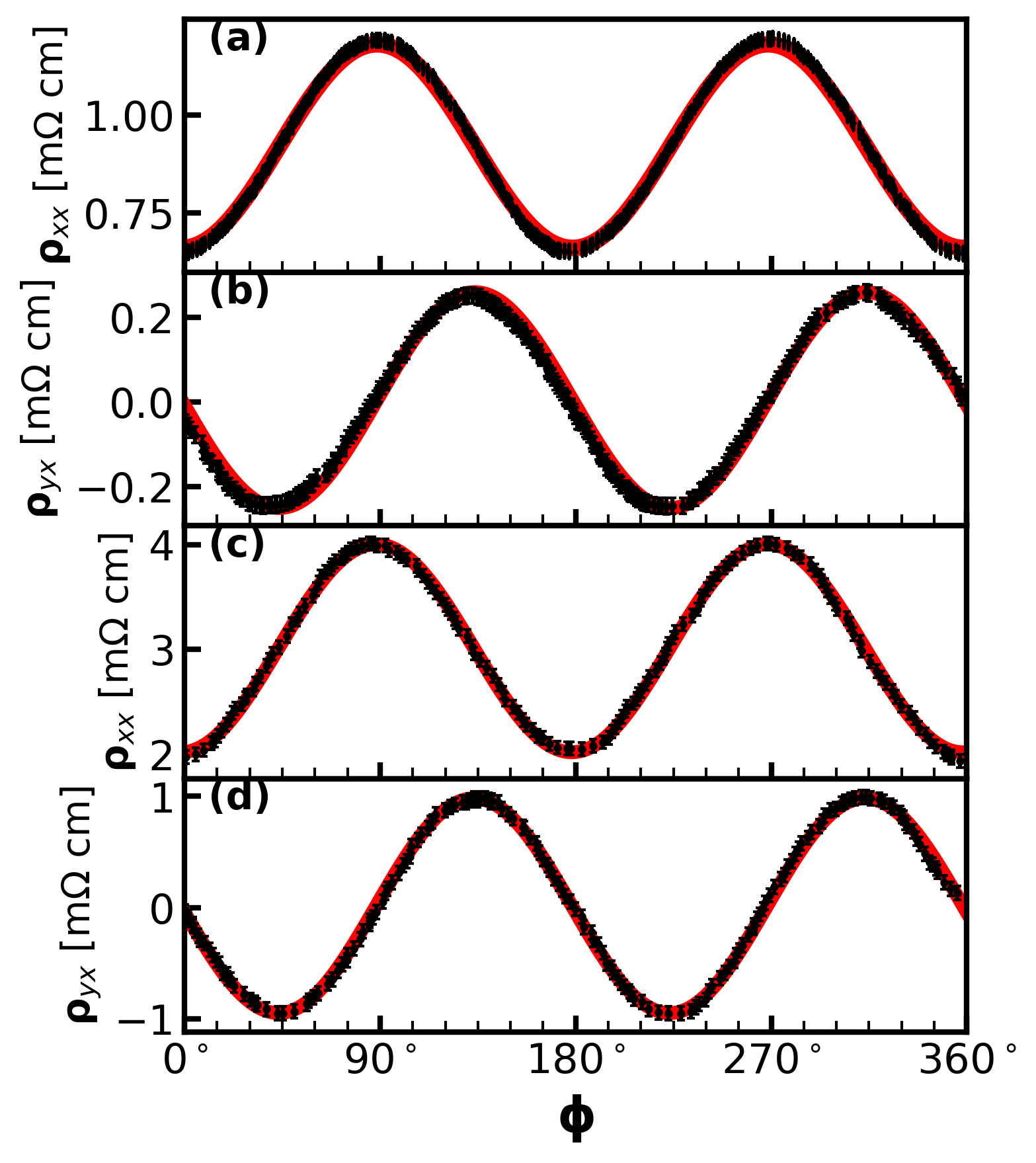}	 
	\end{center}
	\caption{Digitally extracted experimental data for longitudinal and transverse resistivities. Panels (a,b) are taken from Ref.~\cite{Wu2018}, while panels (c,d) are taken from Ref.~\cite{Li2018magnetoresistance}. The black dots represent the digitized experimental data, and the solid lines correspond to fits obtained using our theoretical expressions [Eqs.~(\ref{Resistivity1})and (\ref{Resistivity2})]. The corresponding fitting parameters are listed in rows 1 and 2 of Table~\ref{table1}. The black error bars indicate the uncertainty associated with the digitization process.}
	\label{DataExperiment}
\end{figure}

In Fig.~\ref{DataExperiment}, we digitize the reported \cite{Wu2018,Li2018magnetoresistance} angular-dependent resistivities and fit using   Eqs.~(\ref{Resistivity1}) and (\ref{Resistivity2}), including a constant, $\rho'_0$, offset in the Hall resistivity arising from a small out-of-plane component of the magnetic field. The fitted parameters, namely, $\rho_0$, $\rho_r$, $\phi_r$ and $\rho'_0$ are shown in Table~\ref{table1} for both longitudinal and Hall resistivities in Cd$_3$As$_2$ and trigonal-PtBi$_2$ \cite{ourfootnote}. We note that the parameters $\rho_r$ and $\phi_r$ obtained from fitting $\rho_{xx}$ and $\rho_{yx}$ are almost identical, as proposed, for a given magnetic field. These parameter's weak dependence on magnetic field suggests, the higher order ($B^{2k}, k>1)$ contribution of magnetic field will not alter the angular dependencies on the resistivities. The simultaneous agreement of the amplitude and phase extracted from longitudinal and Hall responses provides a direct test of the proposed phase-shifted form.

\begin{table}[h]
	\caption{Fitting parameters of the digitally extracted data from Refs. \cite{Wu2018,Li2018magnetoresistance,Veyrat2024} in the form of Eqs.\eqref{Resistivity} and \eqref{Resistivity2}, with an additional constant to Hall resistivity. The fitted parameters $\rho_0$, $\rho_r$, and $\rho_0'$ inherit the units of the extracted experimental data reported in the corresponding reference. Consequently, their magnitudes are meaningful only within a given dataset and should not be directly compared between different materials.}
	\begin{tabular}{|c|c|c|c|c|c|c|c|}\hline
	Systems &	B & \multicolumn{3}{c|}{$\rho_{xx}$} & \multicolumn{3}{c|}{$\rho_{yx}$}    \\
		\cline{3-8}
	 & [T]	& \multicolumn{3}{c|}{$\rho_0 -\rho_r \cos 2(\phi +\phi_r)$} & \multicolumn{3}{c|}{$ \rho_0' -\rho_r \sin 2(\phi +\phi_r)$}\\
		\cline{3-8}
 &	& $\,\,\,\,\, \rho_{0}\,\,\,\,\, $ & $\,\,\,\,\, \rho_{r}\,\,\,\,\, $ & $\phi_r$ & $\,\,\,\,\, \rho_{0}'\,\,\,\,\,$ & $\,\,\,\,\, \rho_{r}\,\,\,\,\, $ & $\phi_r$ \\
		\hline 
 	& & & & & & & \\
Cd$_3$As$_2$ \cite{Wu2018} 	&	14 & 0.92 & 0.26 & $1.40^\circ$ & 0.01 & 0.258 & $1.39^\circ$ \\
& 	& & & & & & \\
Cd$_3$As$_2$ \cite{Li2018magnetoresistance}   &	5 & 3.01 & 0.992 & $1.78^\circ$ & 0.02 & 0.987 & $1.78^\circ$ \\		
&	& & & & & & \\
PtBi$_2$\cite{Veyrat2024} 	&	5 & 1.82 & 0.034 & $37.51^\circ$ & -0.14 & 0.035 & $37.17^\circ$ \\	
& 	& & & & & & \\
& 	6 & 1.86 & 0.043 & $35.21^\circ$ & -0.15 & 0.041 & $35.16^\circ$ \\
&	& & & & & & \\  
& 	7 & 1.89 & 0.053 & $34.13^\circ$ & -0.15 & 0.050 & $34.17^\circ$ \\
&	& & & & & & \\
& 	10 & 2.03 & 0.083 & $33.43^\circ$ & -0.16 & 0.076 & $33.35^\circ$ \\
		\hline
	\end{tabular}\label{table1}	
\end{table}

We now estimate the parameters $\alpha_1$ and $\alpha_2$ which determine the parameters $\rho_r$ and $\phi_r$ in diagrammatic Kubo formula approach. We  begin with the minimal low-energy  Hamiltonian for a pair of Weyl nodes with broken time reversal symmetry \cite{NIELSEN1983389,Sharma2016}:
\begin{equation}
	H_\gamma=\gamma v ({\bf k}-\gamma k_{0}\hat{z})\cdot \bm{\tau} \, ,
\end{equation}
where $\gamma =\pm 1$ represents the chirality of Weyl nodes in WSMs, $v $ is the velocity, ${\bf k}$ is the momentum of electrons, and $\bm{\tau}$ are Pauli matrices. The  pairs of nodes with opposite chirality  that respectively act as monopole and anti-monopole \cite{NIELSEN1981219} of Berry curvature in momentum space separated by $2k_0$. For simplicity, we consider a pair of nodes separated along the $z$-direction \cite{PhysRevLett.108.140405}; the qualitative conclusions do not depend on the choice of axis. The corresponding inverse of the Matsubara Green's function for a node of chirality $\gamma$ is given by
\begin{align}\label{greendis}
	\mathcal{G}^{-1}_\gamma({\bf k},i\omega_{n})&=(i\omega_{n}-\mu ) \tau_{0}  -H_\gamma - \Sigma ({\mathbf k},i\omega_n)\, ,
\end{align} 
%=============
%=============
where $\omega_n$ is the fermionic Matsubara frequency, $\mu$ is the chemical potential, $\tau_0$ is the $2\times 2$ identity matrix, and $\Sigma({\bf k}, i\omega_n)$ is the electron self-energy due to scattering from disorder potential. We incorporate elastic disorder through momentum-independent self-energy approximation: ${\bf \Sigma}(i\omega_n \to \omega +i0^{\pm})=\pm ( i/\tau) \tau_0$, where $\tau$  denotes the relaxation time  associated with the  scattering.
%================================
%================================

\begin{figure}
	\begin{center}
			\includegraphics[scale=0.6]{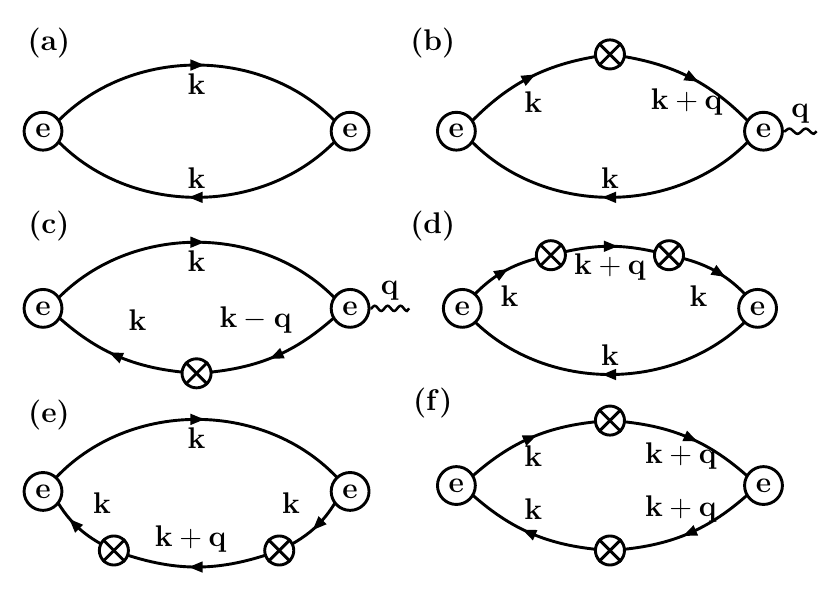}
	\end{center}
	\caption{Feynman diagrams representing Kubo formula. Electric and magnetic vertices for the coupling to electric and magnetic fields are respectively denoted by \textcircled{$\bm e$} and $\otimes$; momenta of fermions are also shown. (a) Zeroth-order diagram in the absence of magnetic field, yielding the Drude and topological Hall conductivities.(b,c) First-order diagrams containing a single magnetic vertex (linear in  $B$). (d,f) Second-order diagrams with two magnetic vertices on the same fermion line. (f) Second-order diagram with the magnetic vertices on opposite fermion lines.}
	\label{Feynman}
\end{figure}

The dc conductivity tensor is obtained from the retarded current-current correlation function, $\sigma_{\mu\nu}=  \text{Im} \left[ \lim_{\Omega \to 0 } \frac{ \Pi^R_{\mu\nu}(\Omega)}{\Omega}\right]$. In the limit of vanishing magnetic field, $\Pi_{\mu\nu}(i\Omega_m)$ in linear response theory is represented by the  Feynman Diagram (Fig.\ref{Feynman}a). This provides the diagonal conductivity $\sigma_{xx} = \sigma_0$ and an anomalous Hall conductivity  $\sigma_{_H} = e^2k_0/(2\pi^2)$,  proportional to the momentum separation between two Weyl nodes with opposite chiralities. 

The current density operators are $\hat{{\bf j}}_\gamma = \partial H_\gamma /\partial {\rm k} =\gamma v \bm{\tau}$. The coupling to the magnetic field enters through the contribution of magnetic vertices given by $e\bm{A}\cdot \hat{{\bf j}}_\gamma$, where $\bm{A}$ is the vector potential due to the application of magnetic field. We choose the gauge $\bm{A} = A_z \hat{z}$ that produces in-plane magnetic fields $\bm{B} = i(q_y  \hat{x} - q_x \hat{y})A_z$, but no out-of-plane component of magnetic field. The linear magnetic field dependent contribution to the dc conductivity does not exist because the contribution of the Feynman diagrams (Fig.\ref{Feynman}b and \ref{Feynman}c) corresponding to a single magnetic vertex identically vanish as the corresponding traces in Eq.\eqref{correlationb2} vanish. The contributions of diagrams in Figs.\ref{Feynman}d and \ref{Feynman}e, wherein both the magnetic vertices in the same fermion line, to dc conductivity cancel each other because the residues in the upper-half plane for $({\cal  G}_\lambda^R)^3{\cal G}_\lambda^A$ and $({\cal G}_\lambda^A)^3{\cal G}_\lambda^R$ are exactly opposite in sign, where ${\cal G}_\lambda^{R,A}$ are retarded and advanced Greens functions. The leading nonzero contribution arises at the quadratic order in magnetic field and corresponds to the Feynman diagram (Fig.\ref{Feynman}f) containing two magnetic vertices in two different fermion lines. The current-current correlation function corresponding to this diagram becomes
{\small \begin{eqnarray}\label{correlationb2}
	&&	\Pi_{\mu
		\nu}= e^{4}v^{4} T \sum_{n,\gamma} \int \frac{d^{3}k}{(2\pi)^3} \text{Tr}\Big[ \tau_{\mu} \mathcal{G}_\gamma ({\bf k},i\omega_{n}+i\Omega_m)(\bm{A}\cdot\bm{\tau})  \nonumber \\
	&& \times \mathcal{G}_\gamma ({\bf k+q},i\omega_{n}+i\Omega_m)\tau_{\nu} \mathcal{G}_\gamma ({\bf k+q},i\omega_{n})   (\bm{A}\cdot\bm{\tau})\mathcal{G}_\gamma ({\bf k},i\omega_{n})\Big]\nonumber  \\ 
\end{eqnarray}
} where the factor $e^4v^4$ is  for four vertices: two electric and two magnetic.
The two magnetic vertices generate both longitudinal and transverse quadratic responses, producing the coefficients $\alpha_1$ and $\alpha_2$ in Eqs.~\eqref{sigma_yx} and \eqref{sigma_xx}. We neglect vertex corrections associated with impurity scattering. Their inclusion may modify numerical coefficients but does not change the symmetry-allowed angular structure derived below.

Expanding the correlation function to second order in the vector potential and leading order in momentum transfer $q$, we obtain
\begin{eqnarray}
	\Pi_{yx}(i\Omega_m,{\bf q}) &\simeq& \Pi_1A_z^2(q_xq_y) + \frac{\Pi_2}{2} A_z^2(q_x^2-q_y^2)   \\
	\Pi_{xx}(i\Omega_m,{\bf q}) & 
	\simeq & \frac{\Pi_1}{2}A_z^2(q_x^2-q_y^2) -\Pi_2  A_z^2(q_xq_y) 
, \end{eqnarray}\color{black}
where form factors
\begin{eqnarray}
	\Pi_1 &=& e^4v^4T\sum_{n,\gamma} \int \frac{k_\parallel dk_\parallel dk_z}{(2\pi)^2} \frac{4v^2(v^2(k_z-\gamma k_0)^2 - \tilde{Z}Z)}{(Z^2-v^2k^2)^2(\tilde{Z}^2-v^2k^2)^2 } \nonumber \\
	\Pi_2 &=& e^4v^4T\sum_{n,\gamma} \int \frac{k_\parallel dk_\parallel dk_z}{(2\pi)^2} \frac{4iv^3 (k_z-\gamma k_0) (\tilde{Z}-Z)}{(Z^2-v^2k^2)^2(\tilde{Z}^2-v
		^2k^2)^2 } \, \nonumber \\ \
		\label{Response_function}
\end{eqnarray}
with $Z = i\omega_n +\mu -\Sigma$ and $\tilde{Z} = i\omega_n + i\Omega_m +\mu -\Sigma$. The momentum integration is performed in cylindrical coordinates with  momentum $k_\parallel$ measured perpendicular to the node-separation axis.  We thus find planar dc conductivities (contributions of Figs.\ref{Feynman}a and \ref{Feynman}f):
\begin{eqnarray}
	\sigma_{yx} &=& 2\alpha_1 B_xB_y + \alpha_2 (B_x^2-B_y^2)   \label{sigma_yx} \\
	\sigma_{xx} &=& \sigma_0 + \alpha_1(B_x^2-B_y^2) - 2\alpha_2 B_xB_y \label{sigma_xx}
\end{eqnarray}
where (see end matter for details)
\begin{eqnarray}
	\alpha_1 &\approx&  
	\frac{ e^{4}}{2(2\pi)^2 k_0^3} \Bigg[\tilde{\tau}^3 \Bigg (\frac{35}{36} -\frac{1}{12\tilde{\mu}^2} +
	\frac{11}{12}\ln \frac{\tilde{\Lambda}}{\tilde{\mu}}\Bigg) \nonumber \\
	&& + \tilde{\tau} \Big( \frac{13}{6\tilde{\mu}^2} + \frac{1}{3\tilde{\Lambda}^2} \Big) +\frac{7}{2\tilde{\mu}^4 \tilde{\tau}} + \frac{3}{4\tilde{\mu}^6\tilde{\tau}^3 }\Bigg] \\
	\alpha_2 &\approx& \frac{e^4 }{(2\pi)^2   k_{0}^3\tilde{\mu}^4} 
\end{eqnarray}
with $\tilde{\tau} = \tau v k_0$, $\tilde{\mu} = \mu/(vk_0)\lesssim 1 $, $\mu \tau >>1$, and $\tilde{\Lambda} = \Lambda/k_0 >>1$, where $\Lambda = \pi/c$ (lattice constant $c$) is the cut-off in transverse momentum \cite{Altland2010Condensed,Johan2018}, {\it i.e.,} $\vert k_z \vert \leq \Lambda$.

\begin{figure}
	\begin{center}
		\includegraphics[width=0.5\textwidth]{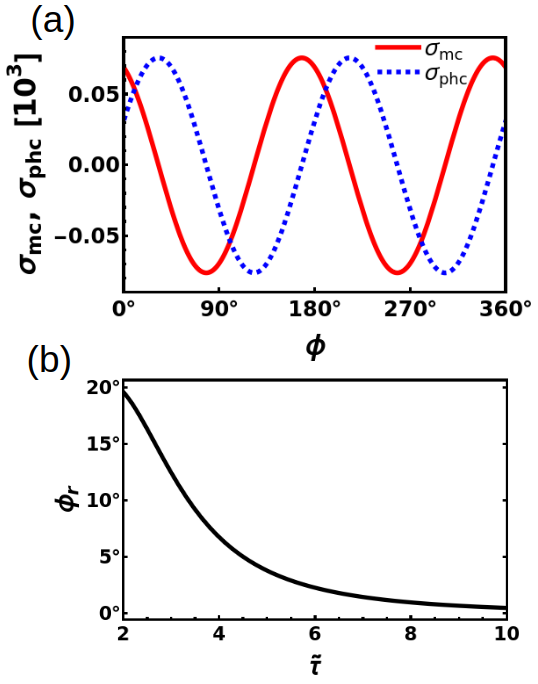}
\caption{(a) Angular dependence of the planar magneto-conductivity $\sigma_{\mathrm{mc}}$ (solid line) and Hall conductivity	$\sigma_{\mathrm{phc}}$ (dashed line) for $\bar{\tau}=3$. Both conductivities are expressed in unit of$\frac{e^4 B^2}{(2\pi)^2 \hbar^3 k_0^3}$	which approximately corresponds to 14 $S/m$ for $B=1$ Tesla and $k_0 = 10^8\,\mathrm{m^{-1}}$. (b) Variation of the phase-shift $\phi_r$ as a function of $\bar{\tau}$.  $\tilde{\mu}=0.5$ and $\tilde{\Lambda}=10$ are considered for calculation.}
\label{Conductivity}
	\end{center}	
\end{figure}

We emphasize that the conventional semiclassical Boltzmann transport theory in the relaxation-time approximation captures only the Fermi-surface contribution to $\alpha_1$, while the intrinsic term $\alpha_2$ arises from the full quantum response.  The coefficient $\alpha_2$ is independent of the disorder relaxation $\tau$, indicating an intrinsic contribution that survives in the clean limit. This appears when the chemical potential $\mu >0$, {\it i.e.}, when $\mu$ is above the energies of the Weyl nodes. Unlike the semiclassical result where $\alpha_1$ scales linearly with $\tau$, the quantum response contains several powers of $\tau$ due to higher-order poles in the response functions \eqref{Response_function}. In the presence of an external magnetic field, the vector potential modifies the electronic wavefunctions and geometric properties of low-energy states near the Fermi surface. Combined with the singular Berry curvature and enhanced orbital magnetic moment near the Weyl nodes, this can generate an intrinsic contribution to the conductivity~\cite{xiao2010berry,nagaosa2010,Burkov2014}. Since transport coefficients receive contributions from all the occupied states below the chemical potential, magnetic-field-induced modifications of these geometric electronic states may play an important role in the observed response. A detailed microscopic understanding of this mechanism remains an interesting direction for future study.

For Weyl semimetals such as TaAs~\cite{Huang2015,Zhang2016}, NbAs~\cite{yang2015}, and TaP~\cite{Arnold2016,Yang2019}, experimentally reported parameters are typically in the range  $\tau \sim 10^{-12}-3\times10^{-12}\,\mathrm{s}$, $\mu \sim 0.02~\mathrm{eV}$, $v \sim (3-4)\times10^5~\mathrm{m/s}$, and $k_0 \sim 10^8-10^9~\mathrm{m}^{-1}$, and  $\Lambda \sim 10^{10}$ m$^{-1}$. Guided by these, we use the dimensionless parameters $\tilde{\mu}=0.5$, $\tilde{\Lambda}=10$, and $\tilde{\tau}=3$ for numerical estimation of the conductivities in the unit of $\frac{e^4B^2}{(2\pi)^2\hbar^3k_0^3}$.  Figure \ref{Conductivity}a shows that the two conductivities acquire identical phase shifts, confirming relation derived in Eqs.~\eqref{sigma_phc} and \eqref{sigma_mc}.Therefore, the same phase shift simultaneously controls the extrema of the planar Hall response and the zero-crossing of the magnetoresistive response. The magnitude of the phase shift, $\phi_r=\frac{1}{2}\arctan(\alpha_2/\alpha_1)$, as a function of the scattering time $\tilde{\tau}$ is shown in Fig.~\ref{Conductivity}b. Since $\phi_r$ depends on the ratio $\alpha_2/\alpha_1$, disorder modifies the phase shift mainly through the disorder-dependent $\alpha_1$. As the strength of disorder increases, $\phi_r$ increases.  The larger phase shift extracted (see Table-\ref{table1}) for PtBi$_2$ may indicate a weaker relative contribution of disorder-dependent transport channels compared with Cd$_3$As$_2$. As per the analysis of experimental data, $\phi_r$ also depends weakly on magnetic field. This is possibly due to the contributions beyond quadratic in magnetic field.

In conclusion, we have shown that planar Hall and magnetoresistive responses of Weyl semimetals acquire a finite phase shift  due to an intrinsic quadratic magnetic-field contribution. Starting from a symmetry-allowed response tensor and evaluation of the corresponding transport coefficients using the Kubo formalism, we find that the angular dependence become $\sin 2(\phi +\phi_r)$ and $\cos 2(\phi+\phi_r)$, respectively. This provides a unified explanation for the displaced planar Hall extrema and angular sign reversal of magnetoresistance observed experimentally. The extracted phase shift therefore serves as a direct probe of intrinsic quantum contributions to magnetotransport in Weyl semimetals.

Acknowledgments:
We thank Arghya Taraphder and Snehasish Nandy for fruitful discussions. S.B.  acknowledges  financial support from   the  Natural  Sciences and Engineering Research  Council of Canada (NSERC).

\bibliographystyle{apsrev4-2}
\bibliography{ref}

\newpage 

\begin{widetext}
		
\section{End Matter}
\noindent {\bf Explicit Calculation of Response Functions $\bm{\Pi_1}$ and $\bm{\Pi_2}$}: We here explicitly evaluate the response functions $\Pi_1$ and $\Pi_2$ in Eq.~\eqref{Response_function}. Fermionic Matsubara frequency sum of a function $F(i\omega_n)$ may be expressed  in terms of a contour integral in a complex plane where residues of the poles at $z= i\omega_n$ equals the frequency sum. The same contour integral may alternatively be evaluated along the contours that turns out to be integrals along parallel to the real axis with just above and just below the real lines along opposite directions. Thus, 	
\begin{align}
	T \sum_n F(i\omega_n) = - \oint_{C} \frac{dz}{2\pi i}\, n_F(z) F(z) 
	= - \int_{-\infty}^{\infty} \frac{d\epsilon}{2\pi i}\, n_F(\epsilon) \big[ F(\epsilon+i0) - F(\epsilon-i0) \big],
\end{align}	
where $\epsilon$ is real. Therefore,
\begin{eqnarray}
	\Pi_{1} &=& e^4 v^4 \sum_\gamma \int \frac{k_\parallel \, dk_\parallel \, dk_z}{(2\pi)^2} 
	\left[ - \int_{-\infty}^{\infty} \frac{d\epsilon}{2\pi i}\, n_F(\epsilon) \left( \mathcal{I}_1^R(\epsilon) - \mathcal{I}_1^A(\epsilon) \right) \right] \\	
	\Pi_{2} &=& e^4 v^4 \sum_\gamma \int \frac{k_\parallel \, dk_\parallel \, dk_z}{(2\pi)^2} 
	\left[ - \int_{-\infty}^{\infty} \frac{d\epsilon}{2\pi i}\, n_F(\epsilon) \left( \mathcal{I}_2^R(\epsilon) - \mathcal{I}_2^A(\epsilon) \right) \right] 
\end{eqnarray}	
where $\mathcal{I}_{1,2}^{R,A} = \mathcal{I}_{1,2}(z=\epsilon \pm i0)$ and
\begin{eqnarray}
	\mathcal{I}_1(z) &=& \frac{4 v^2( v^2 (k_z-\gamma k_0)^2 -\tilde{Z}Z) }{[\tilde{Z}^2 - v^2 k^2]^2 [Z^2 - v^2 k^2]^2}  \\	
	\mathcal{I}_2(z) &=& \frac{4 i  v^3 (k_z-\gamma k_0) \, (\tilde Z-Z)}{[\tilde{Z}^2 - v^2 k^2]^2 [Z^2 - v^2 k^2]^2} 
\end{eqnarray}
with
$Z(z) = z+\mu - \Sigma(z), \qquad \tilde{Z}(z) = z+\mu + i\Omega - \Sigma(z+i\Omega)$.

Assuming short-range scattering potential with energy-independent self energy of the electrons, $\Sigma^{R,A} = \mp i/\tau$. Therefore, the response function $\Pi_2$ ( up to linear in $\Omega$) becomes 
\begin{eqnarray}
	&&	\Pi_2 =\Omega ( 4 e^4 v^7) \sum_\gamma  \int \frac{k_\parallel \, dk_\parallel \, dk_z}{(2\pi)^2} 
	\int_{-\infty}^{\infty} \frac{ (-i) d\epsilon}{2\pi i} n_F(\epsilon)  
	\left[ 
	\frac{(k_z-\gamma k_0)}{[(\epsilon+\mu+\frac{i}{\tau})^2-v^2 k^2]^2}  
	-\frac{(k_z-\gamma k_0)}{[(\epsilon+\mu-\frac{i}{\tau})^2-v^2 k^2]^2 } 
	\right] 
\end{eqnarray}
Because of the fourth-order poles appearing in the correlation function, the evaluation of the Cauchy residues generates contributions involving the Fermi function and its derivatives up to third order with respect to energy, evaluated at the poles. As a result, we obtain
\begin{eqnarray}
	\Pi_2 &=&-\Omega ( 4\, i e^4 v^7) \sum_\gamma  \int \frac{k_\parallel \, dk_\parallel \, dk_z}{(2\pi)^2}  (k_z-\gamma k_0) \Big[\frac{5}{32v^7 k^7}\Big(\Theta (\mu -vk) - \Theta (\mu + vk) \Big) \nonumber \\
	&&+\frac{15}{96v^6k^6}\delta(\mu -vk)
	+\frac{3}{48v^5k^5}\delta^{(1)}(\mu -vk)
	+\frac{1}{96v^4k^4}\delta^{(2)}(\mu -vk) \Big].
\end{eqnarray}
at zero temperature, where $\delta^{(n)}(\mu -vk)$ denotes $n$-th order derivative of $\delta(\mu -vk)$. Since the heavy-side function $\Theta (\mu -vk)$ is nonzero for $\mu > vk$, it restricts the upper bound of the integration in  $k_\parallel$ as $\sqrt{\frac{\mu^2}{v^2} -(k_z-\gamma k_0)^2}$. It further restricts the integration range in $k_z$ as $-\frac{\mu}{v} \leq (k_z-\gamma k_0) \leq \frac{\mu}{v}$. These conditions are also applicable to $\delta$-function term and its derivatives. For the term containing $\Theta (\mu+vk)$, however, no restriction in the integration in $k_\parallel$ is required. The limit in $k_z$ integration is bounded by the inverse lattice spacing: $-\Lambda \leq k_z \leq \Lambda$ with $\Lambda = \pi/c$, wither $c$ is the lattice spacing along perpendicular axis.  With these considerations, the above expression is simplified to
\begin{eqnarray}
	\Pi_2 &=&-\Omega ( 4\,i e^4 v^7) \sum_\gamma \Big[ \int_{-\mu/v}^{\mu/v} \frac{ dk_z}{(2\pi)^2}  (k_z-\gamma k_0)  \Big[  \frac{1}{32 \alpha^7} \Big(\frac{1}{|k_z-\gamma k_0|^5}-\frac{v^5}{\mu^5}\Big) +\frac{ 15}{96v^2 \mu^5} + \frac{1}{2^2 v^2 \mu^5} +\frac{1}{2^3 v^2 \mu^5} \Big] \\&& -\int_{-\Lambda}^{\Lambda} \frac{ dk_z}{(2\pi)^2} \frac{(k_z-\gamma k_0)}{32 v^7 |k_z-\gamma k_0| ^5}\Big]\\ &&=
	- \frac{ i \Omega e^4}{48 (2\pi)^2}  \Big[-\frac{1}{ (\Lambda +k_0)^3}+\frac{1}{ | \Lambda -k_0| ^3} + \frac{1}{\left| \frac{\mu }{v }-k_0\right| ^3}-\frac{1}{\left(\frac{\mu }{v }+k_0\right)^3} -\frac{96k_0 v^4}{ \mu^4}\Big] 
\end{eqnarray}
The integral containing the term $\frac{(k_z-\gamma k_0)}{ |k_z-\gamma k_0| ^5}$ appears to be divergent near the Weyl node at $k_z=\gamma k_0$. However, this divergence is removable in the sense of the Cauchy principal value. The integrand is antisymmetric about the nodal point, and therefore the contributions from the regions $k_z<\gamma k_0$ and $k_z>\gamma k_0$ cancel each other. As a result, although the integrand is singular locally around the node, the total integral remains finite. In the limit $\Lambda >>k_0 >> \mu/v$, only the last term contributes the most. We thus find 
\begin{equation}
	\alpha_2 = \frac{1}{2}\lim_{\Omega \to 0} \frac{\Im \Pi_2}{\Omega} = \frac{e^4 v^4 k_0}{(2\pi)^2 \mu^4} = \frac{e^4}{(2\pi)^2 k_0^3 \tilde{\mu}^4}
\end{equation}

Conductivity depends on the imaginary part of the retarded correlation functions. Since $\Omega$-independent term in $\Pi_1$ is real, we ignore $\Omega^0$ order in $\Pi_1$. The contribution of $\Pi_1$ in linear $\Omega$ is given by 

\begin{eqnarray}
	\Pi_1 
	 && \approx   - 4e^{4}v^{6}\Omega \sum_\gamma \int \frac{k_\parallel \, dk_\parallel \, dk_z}{(2\pi)^2}  \int_{-\infty}^{\infty} \frac{    d\epsilon}{2 \pi i} \frac{R(\epsilon)n_F(\epsilon)}{[(\epsilon+\mu-\frac{i}{\tau})^2-v^2 k^2]^2 [(\epsilon+\mu+\frac{i}{\tau})^2-v^2 k^2]^2} 
\end{eqnarray}	
where
\begin{eqnarray}
	R(\epsilon) &&=-\frac{16  (\mu +\epsilon )}{\tau ^8} \Big[\tau ^2 \Big(v ^6 k^6 \tau ^4-v^4 k^4 \tau ^2 \left(\tau ^2 \left(4 v ^2 (k_z-\gamma k_0)^2+7 (\mu +\epsilon )^2\right)+1\right) \nonumber\\ &&+v ^2 k^2 \left(\tau ^2 \left(8 v ^2 (k_z-\gamma k_0)^2 \left(2 \tau ^2 (\mu +\epsilon )^2-1\right)  +3 \tau ^2 (\mu +\epsilon )^4-2 (\mu +\epsilon )^2\right)-5\right) \nonumber\\ && +4 v ^2 (k_z-\gamma k_0)^2 \left(-3 \tau ^4 (\mu +\epsilon )^4+4 \tau ^2 (\mu +\epsilon )^2-1\right)+3 \tau ^4 (\mu +\epsilon )^6+11 \tau ^2 (\mu +\epsilon )^4+5 (\mu +\epsilon )^2\Big)-3\Big]
\end{eqnarray}
%==================================
The integration contour encloses two poles in the upper half-plane located at $\epsilon_{1,2}= \frac{i}{\tau} \mp v k-\mu,$ and two poles in the lower half-plane located at $\epsilon_{3,4}= -\frac{i}{\tau} \mp v k-\mu$.  Following the same procedure used in the evaluation of $\Pi_1$, we obtain
\begin{align}
	\Pi_{1} 	&=    i \Omega\frac{ e^4 v^6}{(2\pi)^2} \Big[ \frac{11 \tau^3 }{24 v^3  } \log \frac{\Lambda ^2-k_0^2}{(\mu/v)^2-k_0^2}+\frac{\tau}{6 v ^5 } \left(-\frac{1}{\left(\mu/v +k_0\right)^2}-\frac{1}{\left( \mu/v-k_0\right)^2}+\frac{1}{(\Lambda -k_0)^2}+\frac{1}{(\Lambda +k_0)^2}\right) \nonumber\\ & +\frac{\tau^3 \left(\tau^6 \left(35 \mu ^6-3 v ^2 k_0^2 \mu ^4\right)+9 \tau^4 \left(v ^2 k_0^2 \mu ^2+10 \mu ^4\right)+126 \mu ^2 \tau^2+27\right)}{36 v^3  \left(\mu ^2 \tau^2+1\right)^3} \Big]
\end{align}
In the weak-disorder regime relevant to Weyl semimetals, the dimensionless parameter typically satisfies $\mu\tau \gg 1$, while the ultraviolet cutoff obeys $\Lambda \gg k_0$. Under these conditions, the above expression simplifies considerably and can be conveniently expressed in terms of the normalized dimensionless variables as
\begin{align}
	\Pi_{1} &  \approxeq i\Omega \frac{ e^4 }{(2\pi)^2 k_0^3} \Big[ \tilde{\tau}^3 \left(\frac{35}{36}-\frac{1 }{12\tilde{\mu} ^2}+\frac{11  }{12   } \ln \frac{\tilde{\Lambda}}{\tilde{\mu}}\right)+\tilde{\tau} \left(\frac{13}{6\tilde{\mu}^2}+\frac{1}{3\tilde{\Lambda}^2}\right)+\frac{7}{2  \tilde{\mu} ^4 \tilde{\tau}} +\frac{3}{4   \tilde{\mu} ^6 \tilde{\tau}^3}  \Big]
\end{align} 
Therefore
\begin{eqnarray}
	\alpha_1 &=& \frac{1}{2}\lim_{\Omega \to 0} \frac{\Im \Pi_1}{\Omega} \approx  
	\frac{ e^{4}}{2(2\pi)^2 k_0^3} \Bigg[\tilde{\tau}^3 \Bigg (\frac{35}{36} -\frac{1}{12\tilde{\mu}^2} +
	\frac{11}{12}\ln \frac{\tilde{\Lambda}}{\tilde{\mu}} \Bigg) 
	 + \tilde{\tau} \Big( \frac{13}{6\tilde{\mu}^2} + \frac{1}{3\tilde{\Lambda}^2} \Big) +\frac{7}{2\tilde{\mu}^4 \tilde{\tau}} + \frac{3}{4\tilde{\mu}^6\tilde{\tau}^3 }\Bigg] 
\end{eqnarray}
\end{widetext}

\end{document}